\documentclass{pnastwo}

\usepackage{graphicx}
\usepackage{epstopdf}

\usepackage{pnastwof}
\usepackage{amssymb,amsfonts,amsmath}

\def\en{\end{equation}}
\def\p{\partial}

\def\bea{\begin{eqnarray}}
\def\ena{\end{eqnarray}}
\def\p{\partial}

\def\eq#1{{Eq.(\ref{#1})}}
\def\kT{k_{{}_{\rm B}}T}

\begin{document}

\conflictofinterest{The authors declare no conflict of interest}

\track{This paper was submitted directly
to the PNAS office.}

\title{Kinetic regulation of coated vesicle secretion}

\author{Lionel Foret \affil{1}
{Max-Planck-Institute for the Physics of Complex Systems, Nothnitzer Strasse 38, 01187
Dresden, Germany}
\and
Pierre Sens\affil{2}
{ESPCI, rue Vaucquelin, 75005 Paris}}

\contributor{Submitted to Proceedings of the National Academy of
Sciences
of the United States of America}

\maketitle

\begin{article}

\begin{abstract}

The secretion of vesicles for intracellular transport often rely on the aggregation of specialized membrane-bound proteins into a coat able to curve cell membranes. The nucleation and growth of a protein coat
is a kinetic process that competes with the energy-consuming turnover of coat components between the membrane and the cytosol. We propose a generic
kinetic description of coat assembly and the formation of coated vesicles, and discuss its implication to the dynamics of COP vesicles that traffic within the
Golgi and with the Endoplasmic Reticulum. We show that stationary coats of fixed area emerge from the competition between coat growth and the recycling 
of coat components, in a fashion resembling the treadmilling of cytoskeletal filaments. We further show that the
turnover of coat components allows for a highly sensitive switching mechanism between a quiescent and a vesicle
producing membrane, upon a slowing down of the exchange kinetics. We claim that the existence of this switching behaviour, also triggered by
factors such as the presence of cargo and variation of the membrane mechanical tension, allows for efficient regulation of vesicle secretion. We
propose a model, supported by different experimental observations, in
which vesiculation of secretory membranes is impaired by the energy
consuming desorption of coat proteins, until the presence of cargo or
other factors triggers a dynamical switch into a vesicle producing state.
\end{abstract}



\section{Introduction}

The plasma membrane and the membrane of cell compartments such as the ER
and the Golgi continually produce vesicles for cargo transport. Vesicle
formation
generally involves specific proteins that aggregate
into semi-rigid coats of dimensions in the 100nm range, well visible
by electronic microscopy \cite{Heu01,Kir01,Bar02}. The process of
vesicle formation is now rather well established \cite{Mills04}, and is
sketched Fig.1.
First various cytosolic proteins assemble on the membrane into
elementary coat-building units, called monomers in the following.
The membrane-bound monomers then polymerize into coat structures that
locally bend the membrane and recruit cargo molecules.
As the coat expands, the coated membrane invaginates until
forming a nearly spherical vesicle containing cargo \cite
{Gal05,Ant06}, that is eventually released from the membrane. The
coat components soon disassemble and are ready to participate to the
formation of a new vesicle.

The coats are classified in three major classes, COPII, COPI and
Clathrin. Although they involve distinct proteins, the three types of
coat
share many common features, from their size and shape to the
mechanism by which polymerization, cargo recruitment and
membrane deformation is achieved \cite{Mills04}. Our approach is
primarily aimed at studying the formation of COPI and
COPII vesicles. However, the generality and robustness of its
outcome suggest relevance for the more sophisticated Clathrin coats
as well. The assembly of COPs and Clathrin, and the fission of COP
vesicles can now be reconstituted on purified liposomes with a restricted
number of components \cite{Yeu98,Orc98,Cam98}.
Those experiments point-out the robustness of the coat formation
process. They also confirm that coat polymerization is
spontaneous, only driven by weak short range attractions
between the monomers, while coat disassembly requires the presence
of an energy source. More precisely, the assembly and disassembly of
COP coat components follow the cycle of activation - inactivation of
a GTPase protein, Sar1 for COPII and Arf1 for COPI \cite{Mills04}.
Once activated, the GTPases bind to the membrane and recruit individual
coatomer complexes (the monomers), that later polymerize into coats
\cite{Lip01}. The inactivation of the GTPase, triggered by the
hydrolysis of its bound GTP, leads to its unbinding from the membrane
and to the monomer disassembly if the GTPase belongs to a monomer.

Strikingly, FRAP experiments suggest that the exchange kinetics of  coat
components is much faster than the rate of vesicle
secretion \cite{Lip01}.
In other words, many futile monomers are released to the cytosol
during the expansion of a coat. So, while new membrane-bound
monomers polymerize at the coat periphery, others within the coat
disassemble and are expelled to the
cytosol. Paradoxically, the consumption of energy \textit{via} GTP
hydrolysis
seems to work against coat growth and to prevent vesicles
formation. This resembles microtubules dynamics and by analogy to
the treadmilling of microtubules, it has been suggested that the
competition between growth
and unbinding may produce stable coats of fixed area \cite
{Lip01,Lip03,Lip05,Ung06}.

In this paper, we investigate theoretically the consequence of futile
release of coat components on the distribution of size and shape of
protein coats, and more practically on the amount of secreted
vesicles. In our model (Fig.1), monomers are continuously "dropped"
onto a
membrane and proceed to aggregate into coats of growing size that
curve the membrane. Monomers leave the membrane either individually
after GTP hydrolysis or collectively as part of a completed vesicle.
Intuitively, one expects GTPase inactivation to decrease the rate
of vesicle formation by
reducing the lifetime of membrane-bound monomers. 
However, our generic approach reveals that a deeper
understanding of vesicle secretion requires a quantitative statistical
model.
Indeed, we report the existence of a discontinuous dynamical
transition from a quiescent to a vesicle
producing membrane, upon variation of the rate of GTP hydrolysis. In other words, the apparently counter-productive
energy consumption that favors the unbinding of coat components provides
secretory membranes with a highly sensitive switch to regulate
vesicle release, triggered for instance by a variation of the cargo
concentration or the mechanical tension of the membrane.
\begin{figure}
\vspace{-0.8cm}
\centerline{
\includegraphics[scale=0.38]{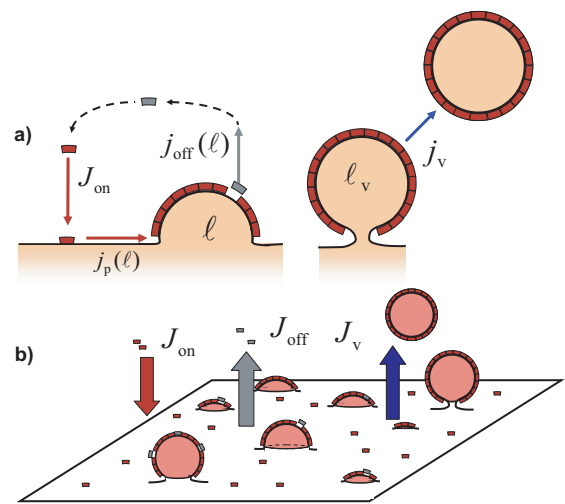}}
\vspace{0.3cm}
\caption{
{\bf a)} A monomer cycle: activation/membrane binding ($J_{\rm on}$), aggregation ($j_{\rm p}(\ell)$), 
membrane unbinding {\textit via} inactivation ($j_{\rm off}(\ell)$) or vesiculation ($j_{\rm v}$). 
{\bf b)} Coat population and the global fluxes in and out of the membrane.}
\end{figure}

\vspace{-0.5cm}\section{Description of the model}

Our goal is to describe the collective behaviour of a population of
evolving membrane domains (coats) formed by the aggregation of identical
units (monomers), which are themselves continuously recycled between the
membrane and a reservoir (the cytosol). Our starting point is the course
of events depicted in Fig.1. We consider a patch of membrane much larger
than the size of individual coats, which is subjected to a constant and
homogeneous in-flux of monomers $J_{\rm{on}}$. The monomers have a
finite
lifetime on the membrane before being recycled to the cytosol, at a
rate $k_{\rm off}$. While on the membrane, they diffuse and eventually aggregate into curved protein coats. Coats that manage to reach a
critical size leave the membrane as coated vesicles.

\underline{\em Monomers}: The formation of a new monomer on the membrane 
involves a succession of steps (GTPase binding on membrane and activation, and the
recruitment of coat proteins) which are not individually described in the present model. 
The rates associated with these processes enter a unique parameter, the mean number of monomers 
$J_{\rm{on}}$ formed on the membrane 
per units of time and area.

\underline{\em Coat growth}: Coat expansion proceeds by polymerization of 
monomers at the coat edge.
Monomer-monomer binding is spontaneous and results from weak short range
interactions. The binding energy $\gamma$ should be in the range of a few $k_{{}_{\rm B}}T$ 
($k_{{}_{\rm B}}T$ is the energy available from thermal fluctuations, 
with $k_{{}_{\rm B}}$ the Boltzmann constant and $T$ the temperature in Kelvin), 
since the $10 k_{{}_{\rm B}}T$ provided by GTP 
hydrolysis is sufficient to break the bonds. 
The polymerization is thus thermally reversible and solely driven by the minimization 
of the free energy of the coat.

\underline{\em Coat structure}: Electron
microscopy \cite{Heu01,Kir01} supports the assumption
that the
optimal area per monomer $s_0$ ($\sim 100\ \text{nm}^2$) and the optimal
radius of curvature of the coat $R_0$ ($ \sim 50\ \text{nm}$) are
homogeneous within the coat and remain constant during coat growth.
We thus adopt a model in which the state of a coat is fully
characterized by a single, slowly varying parameter: the number of
polymerized monomers $\ell$ it contains, taken as a continuous variable
for commodity. All other internal degrees of freedom in the coat
(protein density and coat shape) are considered to adjust to their
optimal
configuration faster than the typical rates of coat growth and GTP-hydrolysis.  Under
these assumptions, a given coat of size $\ell$ can be described as a
spherical cap of constant curvature (defined as the dimensionless
quantity
$c=\sqrt{s_0/(4\pi R_0^2})\sim 1/20$). A full spherical coat ($\ell c^2=1$)
with these properties
contains several hundreds monomers.

\underline{\em Monomer release}: Contrary to the reversible
monomer polymerization, monomer desorption is an energy consuming
process driven by GTP hydrolysis. It occurs at a rate $k_{\rm{off}}$,
assumed constant here for simplicity (see Supporting Information (SI) for a discussion of this assumption). This rate can be estimated from
FRAP
experiments \cite{Lip01,Lip05,Ant03,Pep06}, $k_{\rm{off}}\sim 0.1-10\
s^{-1}$. Under the assumption of constant protein density in the
coat, the dissociation of monomer from the coat is followed by a rapid
rearrangement of the coat structure and by a slight shrinkage of the coat; monomer
inactivation thus opposes coat expansion.

\underline{\em Vesicle release}: A mature coat containing a number
$\ell_{\rm{v}}$ ($\ell_{\rm v}c^2\lesssim 1$) of bound monomers forms a
nearly closed
sphere connected to the rest of the membrane by a thin neck. At this
stage, the coat cannot grow further, and is eventually released into
the
cytosol as a coated vesicle. The details of the scission mechanism vary
between classes of coat, and may involve additional proteins (\cite{Mills04}).
Here, we merely assume that once a domain reachs the critical size
$\ell=\ell_{\rm{v}}$, it leaves the membrane as vesicles at a
constant rate
$k_{\rm{v}}$.

\underline{\em Membrane properties}: The curvature of the coat imposes a
deformation to the membrane which is opposed by membrane tension
\cite{sens_cav}. The membrane tension $\sigma$ thus favors coat
depolymerization, and can have a sizable effect on coat growth if it is larger than  
$\gamma/(s_0\sqrt{\ell_{\rm v}})\simeq 10^{-5}J/\text{m}^2$ (A detailed model for the elastic properties of the protein-covered membrane is discussed in SI, section II.B).
Tensions of the Golgi, the ER, and the plasma membranes are typical
in the range $\sigma \sim 10^{-6} - 10^{-4} J/ \text{m}^2$ \cite{tension_ref}, 
and may thus play a role in vesicle secretion. Hereafter,
membrane tension will be expressed in natural units:
$\bar{\sigma}=\sigma s_0\sim 10^{-2} - 1 \kT$.

The population of membrane coats is characterized by its size
distribution
$n(\ell)$. The mean concentration of coats of size $\ell$ (between
$\ell$ and $\ell+d\ell$) at a time $t$ is
$n(\ell,t)d\ell$, and the mean concentration of isolated
monomers
is $n_1(t)$.
Our purpose is thus to compute, $n(\ell)$ and $n_1$ at steady state,
for given values of the parameters
$\gamma$, $\bar{\sigma}$, $k_{\rm{off}}$, $J_{\rm{on}}$, $k_{\rm{v}}$ and $
\ell_{\rm{v}}$. More practically, we will compare the fluxes of coat
elements leaving the membrane as inactive monomers $J_{\rm{off}}$ and as part of a vesicle $J_{\rm{v}}$ (Fig.1).


\section{Theoretical framework}

\subsection{Monomer fluxes and conservation relations}
In this section, we derive the kinetic equations for the evolution of the coat size distribution $n(\ell)$. The monomer cycle can be divided into four steps, to which correspond
four different fluxes, as shown Fig.1.

\noindent- $J_{\rm{on}}(t)$ is the in-flux of single monomers binding to
the membrane, taken as an input in our model.\\
- $j_{\rm{p}}(\ell,t)$ is the flux of monomer joining domains of
size $\ell$, a balance between polymerization and
depolymerization for this domain size. Integrated over the entire
population, it gives the total flux of monomers incorporated into
domains $J_{\rm{p}}=\int_1^{\ell_{\rm{v}}} d\ell\ j_{\rm{p}}(\ell,t)$.\\
- $j_{\rm{off}}(\ell,t)$ is the flux of monomers expelled from
domains of size $\ell$ into the
cytosol after GTP hydrolysis. Under the assumption of uniform release, it is given by $j_{\rm{off}}(\ell)=k_{\rm{off}}n(\ell)\ell$. Integrated over the entire population,
it gives the total flux of individual inactive monomers leaving the
membrane $J_{\rm{off}}=\int_1^{\ell_{\rm{v}}} d\ell\ j_{\rm{off}}
(\ell,t)$.\\
- $j_{\rm{v}}(t)$ is the flux of mature coats released as
vesicles. Introducing the rate of vesicle formation $k_{\rm v}$, we have $j_{\rm{v}}=k_{\rm{v}}n(\ell_{\rm{v}})$. The total flux
of monomers leaving the membrane as part of a vesicle is $J_{\rm{v}}(t)=j_{\rm{v}}\ell_{\rm{v}}$.

The evolution of the coat size distribution satisfies (see SI):
\begin{equation}
\p_t n(\ell)=-\p_\ell\left(j_{\rm{p}}(\ell)-j_{\rm{off}}(\ell)\right),\label{conservation1}
\end{equation}
and that the  polymerization current $j_{\rm{p}}$ reads:
\begin{eqnarray}
j_{\rm{p}}(\ell)=-k_{\rm{p}}n_1\big(\partial_\ell n(\ell)+n(\ell)\p_{\ell}\Delta E
(\ell)\big).\label{jp}
\end{eqnarray}
$j_{\rm p}$ is proportional to the density of
available monomers $n_1$, and to the rate of monomer binding onto coats $k_{\rm{p}}$ (see below). It contains a diffusive term accounting for random polymerizations and
depolymerization induced by thermal noise, and  a convective term
describing the drift of monomers toward domains of lower
energy, driven by the ``force'' $-\p_{\ell}\Delta E$. The free energy
difference $\Delta E$ (in $\kT$ units) between a coat of size $\ell$
and $\ell$
isolated monomers  diffusing on the membrane is obtained treating the coat as a rigid spherical cap 
\cite{Sen06} (see also SI):
\begin{eqnarray}
&\Delta E(\ell)=\gamma \sqrt{\ell(1-c^2\ell)}+\bar{\sigma} c^2\ell^2-
\mu(n_1)\ \ell\label{El0}\\
&{\rm where} \qquad \mu=\ln n_1+\gamma\sqrt{1-c^2}+\bar{\sigma} c^2\label{mu}
\end{eqnarray}
is a chemical potential including the entropy of the  freely diffusing monomers, see SI.

The net
current $j(\ell)=j_{\rm{p}}(\ell)-j_{\rm{off}}(\ell)$ accounts for  polymerization and desorption. It  can be
written in terms of an effective energy $\tilde{E}(\ell)\equiv\Delta E(\ell)+\frac{k_{\rm{off}}}{2k_{\rm{p}}n_1}(\ell^2-1)$ :
\begin{equation}
j=-k_{\rm{p}}n_1(\partial_\ell n+n\p_\ell \tilde{E}),\label{j}
\end{equation}
with
\begin{eqnarray}
\tilde{E}(\ell)=\gamma \sqrt{\ell(1-c^2\ell)}+\Sigma(n_1)
\ell^2-\mu(n_1)\ \ell+{\rm const.},\label{El}\\
\Sigma(n_1)\equiv\bar{\sigma} c^2+\frac{k_{\rm{off}}}{2k_{\rm{p}}}\frac{1}{n_1}.\label{sigma}
\end{eqnarray}
This equation introduces an effective tension $\Sigma$ that illustrates the fact that desorption of inactive monomers and membrane tension formally play the same role in  hindering coat maturation and vesicle secretion. Note that more generally, the coatomer binding or inactivation rates may depend on the coat size, in which case monomer desorption  enters the effective energy as $\int d\ell \ell (k_{\rm off}(\ell)/k_{\rm p}(\ell))$.

Finally, the monomer influx $J_{\rm on}$ and the flux of secreted vesicle $j_{\rm v}$ are accounted for via the boundary conditions (see SI)
\begin{eqnarray}
&&j_{\rm{p}}(1)=J_{\rm{on}}-J_{\rm{p}},\label{conservation2}\\
&&j_{\rm{p}}(\ell_{\rm{v}})-j_{\rm{off}}(\ell_{\rm{v}})=j_{\rm{v}},\label{conservation3}
\end{eqnarray}


\subsection{Steady state}
At steady state, all fluxes are balanced and $\partial_t n=0$. 
Eqs.(\ref{conservation1}-\ref{conservation3}) 
reduce to two
conditions to be satisfied by  $n_1$ and $n(\ell)$:
\begin{eqnarray}
j(\ell)&=&j_{\rm{v}}={\rm constant}.\label{jl}\\
J_{\rm{on}}&=&J_{\rm{off}}+J_{\rm{v}},\label{Jon}
\end{eqnarray}
The former equation enforces that the size distribution is constant, and the latter  that  the flux of
monomer binding to the membrane balances the flux of monomer leaving
the membrane, either after
inactivation or by vesiculation.

\section{Results}

In this section, we focus on the steady state of a membrane receiving a constant in-flux of monomer, 
each having a finite lifetime at the membrane. The full characterization of the coat population and 
of vesicle secretion follows two steps. First, the stationary distribution of coat size $n(\ell)$ 
is computed for a given concentration of free monomers $n_1$ with \eq{jl}.  Second,  $n_1$ is self-consistently derived for a given  monomer in-flux $J_{\rm on}$ by imposing that the in-flux matches the total monomer out-flux (\eq{Jon}).
While the first step relies entirely on the properties of the free energy landscape $\tilde{E}(\ell)$, 
the second introduces collective effects emerging from the competitive growth of many domains, 
which ultimately give rise to the ``secretory switch''.\\

\subsection{Effective energy landscape and steady-state distribution}

Apart from thermal fluctuations, a coat is driven toward growth or shrinkage by the effective ``force''
$-\p_{\ell}\tilde{E}(\ell)$ (\eq{j}). Coat growth is thus formally analogue to thermal diffusion 
along the effective energy landscape $\tilde{E}(\ell)$ (\eq{El}). The analytical expression of the coat size distribution is given in the SI. Several different regimes can be distinguished under increasing monomer concentration (Fig.2). 

The energy landscape illustrates the interplay 
between antagonistic effects; short-range attractions between monomers promotes 
polymerization, while the entropy of the free monomer favors their dispersion. The $\gamma$ and $\mu$ terms in \eq{El} reflect this competition. Furthermore, monomer inactivation and unbinding, and  
membrane mechanical tension hinder coat growth ($k_{\rm off}$ and $\sigma$ respectively, combined in 
the effective tension $\Sigma$, \eq{sigma}). Depending on the monomer concentration, the landscape may show a local maximum at a small size $\ell_{\rm n}$  (Fig.2\textbf{b}-\textbf{d}), which indicates a nucleation process. Coats must reach the critical size $\ell_{\rm n}$ (through fluctuations in the pool of free monomers) in order to consistently grow further, and the rate of nucleation is controlled by the height of the energy barrier. The effective energy may also show an barrier to vesiculation for large coat size $\ell\sim\ell_{\rm v}$ (Fig.2\textbf{b}-\textbf{c}), indicating that large coats are suppressed by the effective membrane tension $\Sigma$. A local minimum then exist for an intermediate size $\ell^*$, corresponding  to kinetically stable coats.

\begin{figure}
\vspace{-0.3cm}
\centerline{
\includegraphics[scale=0.4]{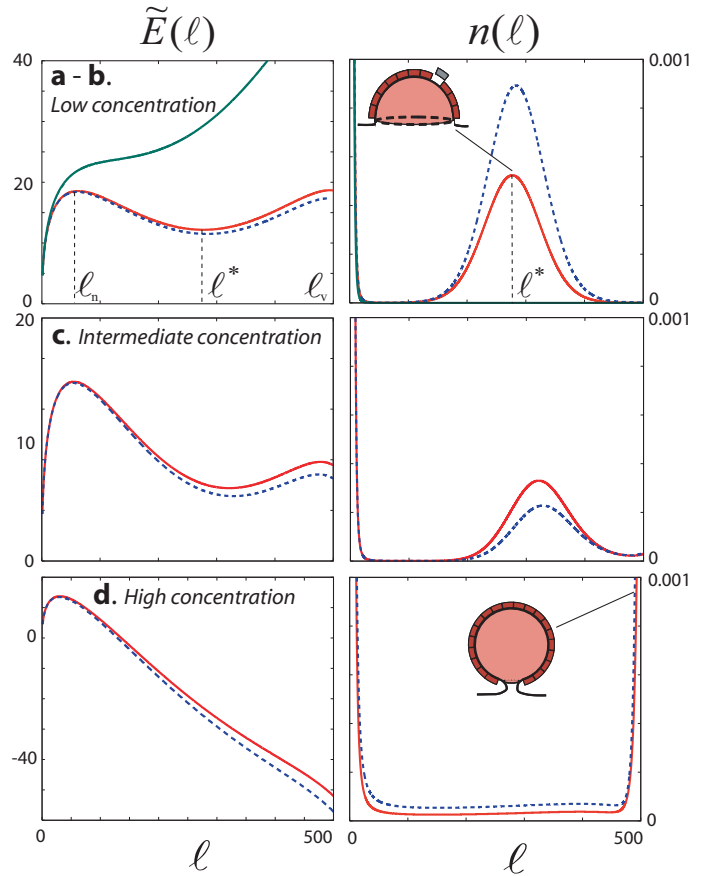}}
\caption{Effective energy landscape $\tilde{E}(\ell)$ (left column, in $\kT$ units, from \eq{El}) 
and the corresponding coat size distribution $n(\ell)$ (right column, in $n_1^{\rm{cmc}}$ units, 
from Eqs.(\ref{j}-\ref{jl})), for different values of the free monomer density 
$n_1$. $n_1/n_1^{\rm{cmc}}=0.986$ ({\bf a} - first row green curve), $1.039 $ ({\bf b} - first row red),  $1.054$ ({\bf c} - second row) and $1.160$ ( {\bf d} - third row). 
The variation of $\tilde{E}(\ell)$ and $n(\ell)$ upon a slight increase of 
$n_1$ above the given value are shown in blue dashed lines. 
Other parameters are $\gamma=5\ k_{{}_{\rm{B}}}T$, $\sigma=0$, $k_{\rm{v}}=0.13 s^{-1}$, 
$k_{\rm{off}}=0.12 s^{-1}$ 
and, $\ell_{\rm{v}}=500$. With those values, $n_1^{\rm{cmc}}=0.009$.}
\end{figure}

\noindent{\bf Low monomer concentration}. At low monomer concentration, the large energy barrier to vesiculation at $\ell=\ell_{\rm v}$ prevents coats to mature into vesicles ($J_{\rm{v}}\simeq 0$)   (Fig.2, states \textbf{a} and \textbf{b}).  The coat size distribution resembles a distribution at thermal equilibrium: 
$n(\ell) \simeq e^{-\tilde{E}(\ell)}$, and the membrane follows a classical scheme common to many self-assembling systems ({\em e.g} surfactants in solution \cite{isra}). The local energy minimum at $\ell^*$ appears above a critical concentration  $n_1^{\rm cmc}$, analogue to the  ``critical micellar concentration'', or ``cmc'' at which surfactants in solution start forming micellar aggregates (see \cite{isra} and the SI). For $n_1<n_1^{\rm cmc}$ (Fig.2{\bf a}), entropy dominates and the effective energy increases  monotonously with the coat size $\ell$. Monomer aggregation is unfavorable, and the membrane contains mainly single monomers and few small transient domains formed by fluctuation. For $n_1>n_1^{\rm cmc}$ (Fig.2{\bf b}), long-lived coats can nucleate, at a rate fixed by the nucleation barrier,  and grow up to the optimal size $\ell^*$. Maturation into coated vesicles ($\ell=\ell_{\rm v}$) is prevented by monomer desorption and membrane tension.

\noindent{\bf Larger monomer concentration}. The height of the energy barrier to vesiculation at $\ell=\ell_{\rm v}$ decreases with increasing monomer concentration. When it falls below the nucleation energy barrier (Fig.2{\bf c}), domains may mature into fully-formed vesicles and vesicle secretion becomes increasingly probable. The optimal coat size $\ell^*$ increases with $n_1$, and  eventually exceeds the critical size for vesiculation $\ell_{\rm{v}}$ at high monomer concentration (Fig.2{\bf d}), under which conditions any 
nucleated domain matures into a fully formed vesicle.

The growth of individual coats is controlled by the amount of free, active monomers $n_1$.
On the other hand, the pool of free monomer is depleted by their binding onto growing coats and is thus influenced 
by the coat population. As we shall see next, this feedback induces remarkable collective effects within the coat 
population, which presents a discontinuous transition between a state of arrested growth and a state of 
abundant vesiculation within a narrow range of kinetic parameters. \\


\subsection{Vesicle secretion is controlled by collective effects}  
\begin{figure}
\vspace{-0.5cm}
\centerline{
\includegraphics[scale=0.75]{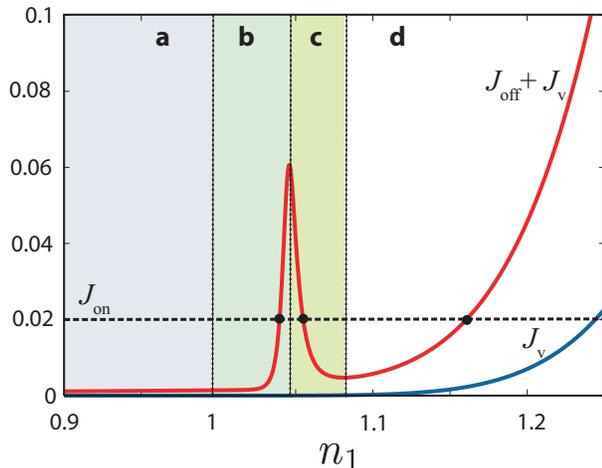}}
\caption{The fluxes of monomers leaving the membrane at steady state, as a
function of the density of active monomer $n_1$ (in
$n_1^{\rm{cmc}}$ unit).  $J_{\rm v}$ (blue) is the vesiculation flux and $J_{\rm off}$ (red) the sum of vesiculation and inactivation. 
At stationary state, the total outgoing flux balances the incoming flux
($J_{\rm off}+J_{\rm v}=J_{\rm on}$) (horizontal dashed line).
Fluxes are in $k_{\rm{p}}(n_1^{\rm{cmc}})^2$ unit and the black dots correspond to the three different states ({\bf b}, {\bf c}, {\bf d}) depicted in Fig.2.}
\end{figure}

The solution of the coupled Eqs.(\ref{jl},\ref{Jon}) is graphically represented on Fig.3 
as the intersection of the monomer in-flux $J_{\rm on}$ and total out-flux $J_{\rm off}+J_{\rm v}$. It may fall in four different regimes ({\bf a} to {\bf d}), corresponding to the four distributions plotted in Fig.2.
In a wide range of parameters (see below),
$J_{\rm{off}}$ displays the remarkable property of being non-monotonous, with a sharp peak at a critical concentration of free monomers. This behavior dramatically influences the membrane's ability to secrete vesicles. Indeed, a given monomer in-flux may correspond to three distinct dynamical states of the membrane. We will show below that regimes {\bf b} and {\bf d} represent respectively a quiescent membrane and a membrane secreting large amount of vesicles, while regime {\bf c} is dynamically unstable. 
The secretory membrane thus constitutes a bistable dynamical system 
able to abruptly switch vesicle secretion on and off at prescribed monomer turnover rates. 

Since all membrane-bound monomers are inactivated with the same rate $k_{\rm off}$, the total flux of monomer leaving the membrane after inactivation $J_{\rm off}$ is directly proportional to the total amount of monomer on the membrane. The peak of $J_{\rm off}$ in Fig.3 stems from the complex relationship between the concentration of isolated monomer $n_1$ and the total amount of coat components on the membrane. 

\noindent{\bf No vesicle secretion regimes a and b}. If no coat can form (low monomer concentration: state {\bf a}), the monomers out-flux is dominated by the desorption of free monomer: $J_{\rm off}\simeq k_{\rm off} n_1$. If coats can form, but do not mature into vesicles (state {\bf b}),  the out-flux is dominated by the desorption of monomers belonging to coats of size $\ell^*$: $J_{\rm off}\simeq k_{\rm off} \ell^* n(\ell^*)$. In this regime, monomers reaching the membrane tend to join a coat and the density of free monomer is almost insensitive to the fluxes: ($n_1\sim n(\ell^*)^{1/\ell^*}$ with $\ell^*\gg1$, see SI). A small increase of $n_1$ requires a pronounced
increase of the total amount of coat material on the membrane, which explains the sharp rise of the total monomer out-flux $J_{\rm off}$ with $n_1$ in Fig.3. In this regime, the optimal coat size is also insensitive to the monomer fluxes, and is obtained from the minimization of the effective energy $\tilde E$ (\eq{El})
\begin{equation}
\ell^{*}\propto\bigg(\frac{\gamma}{\Sigma (n_1^{\rm{cmc}})}\bigg)^{2/3},
\label{lstar}
\end{equation}

\noindent{\bf   The unstable regime c}. For intermediate monomer density, metastable coated pits have a high probability to grow into fully
formed vesicle owing to the small barrier to vesiculation (Fig.2{\bf c}). The rate of vesicle formation increases with $n_1$, so the total amount of membrane-bound material actually {\em decreases} with increasing concentration of free monomer. This is shown by the dashed blue line in Fig.2{\bf c}, and explain the decrease of $J_{\rm off}$ in Fig.3. This situation cannot be maintained at steady state, and spontaneously evolves
toward either state {\bf b} or {\bf d}.

\noindent{\bf Steady vesicle secretion regime d}. If the monomer concentration on the membrane is large, the effective coat energy exhibits a nucleation barrier but 
no intermediate minimum (Fig.2{\bf d}). 
After nucleation, a coat grows at nearly
constant velocity until reaching the critical size $\ell_{\rm{v}}$ where it remains trapped for a time $1/k_{\rm{v}}$ 
before being released as a vesicle. In this regime, both  $J_{\rm off}$ and $J_{\rm v}$ increase with $n_1$, Fig.3.

The bistability exhibited by the coats dynamics relies on the existence of an unstable steady state and holds as long as there exist a (meta)stable 
 coat of intermediate size. This feature is conserved even if the effective energy contains higher order terms, to be expected if the ratio of monomer dissociation 
to binding rates ($k_{\rm off}/k_{\rm p}n_1$) increases with the coat size (see SI). Furthermore, the switch exists if there is a metastable-state within the accessible size-range: $\ell^{*}/\ell_{\rm{v}}(\sim\ell^{*}c^2)<1$.  From \eq{lstar}, this condition amounts to 
$\Sigma>\gamma c^3(\sim 10^{-4})$. The effective tension $\Sigma$ (\eq{sigma}) accounts both for the 
membrane mechanical tension ($\bar{\sigma} c^2\simeq 10^{-5}-10^{-3}$) and the ratio  ($k_{\rm off}/k_{\rm p}n_1$). The binding rate is assumed to be limited by monomer diffusion, and is expected to be of order the inverse monomer  diffusion time over its own size 
($k_{\rm p}\sim D/s_0\sim 10^4 s^{-1}$, with $D\sim \mu m^2/s$ the membrane diffusion coefficient). With a dissociation rate  $k_{\rm off}\simeq 1s^{-1}$, and a monomer density $n_1\sim 1\%$ (or $k_{\rm p} n_1\sim10^{2}s^{-1}$), we find that secretory membranes 
are well into the bistable regime ($\Sigma\simeq 10^{-2}\gg\gamma c^3$), and should exhibit the secretory switch 
discussed below.


\section{Discussion}

The growth of coated pits and the secretion of coated vesicles result
from a kinetic balance between the polymerization and the inactivation of coat
components. It is thus to be expected that coat maturation can only
proceed if the coatomers turnover at the membrane is sufficiently slow \cite{Lip01,Lip03,Ung06}.
Our model goes beyond this intuitive analysis, and shows that secretory
membranes are able to abruptly switch between a quiescent and a vesicle
producing state upon a slowing down of coatomer recycling. Switch-like behaviors are clearly advantageous for biological
systems. The highly non-linear nature of a switch confers evident
robustness with respect to the noisy environment, and allows for a
precise regulation of the system's activity. The biological consequences of the
``secretory switch'' are discussed below, together with the influence of
important factors, such as the density of cargo or the membrane
tension, in regulating the activity of secretory membranes. We also
show how some apparently unrelated observations on COPs vesicles naturally
fit into our global picture of coated vesicle secretion.\\


\subsection{The ``secretory switch''}

Consider a secretory membrane receiving a fixed amount of coatomer per unit time. 
As coatomers accumulate and aggregate, 
the membrane eventually reaches a steady state in which the flux of coatomer leaving the membrane 
(by inactivation or vesicle secretion) balances the in-flux. 
If the in-flux is low, the membrane is covered by monomers and stationary coated pits, 
the latter being prevented to mature into fully-formed coated vesicles by the combined 
effect of coatomer recycling and the energy associated to membrane deformation. 
As the in-flux increases, the coated pits to monomer ratio increases (Fig.2{\bf a}, dashed blue line), 
but as long as the membrane remains in the ``stationary pits" regime 
({\bf b} - Fig.3)  the size and shape of the coated pits is weakly sensitive 
to the in-flux and no vesicle is produced. However, beyond a  threshold 
value of the in-flux, state {\bf b} disappears (Fig.3). 
Coated pits are not kinetically stable, and
after a fast transient regime that sees the release of the previously stable coated pits,  
the membrane settles into a state of steady vesicle secretion (state {\bf d}).

If the in-flux is now reduced, the membrane {\em remains} in the secretory state {\bf d}, 
which possesses a stable branch for smaller in-flux (Fig.3). The secretory regime disappears 
below yet another critical threshold 
and the systems jumps back into the
quiescent state {\bf b}.

The coat population can thus undergo discontinuous dynamical transitions, 
characterized by an hysteretic cycle, between the two non-equilibrium
steady states {\bf b} and {\bf d}. In other words, the secretory membrane works 
in an all-or-nothing fashion and can switch vesicle production
on and off within a narrow range of control parameters such as the 
GTPase activation and inactivation rates, and membrane tension.
Fig.4 shows the rate of vesicle secretion as a function of the GTP hydrolysis rate 
$k_{\rm{off}}$. 
The discontinuous transition between quiescent and vesicle-producing membrane is clearly apparent, 
and is characterized by two hydrolysis rates (high and low thresholds). Between these two rates, 
the secretory membrane may be in either state, depending on the system's history (hysteresis).

\begin{figure}
\vspace{-0.4cm}
\centerline{
\includegraphics[scale=0.55]{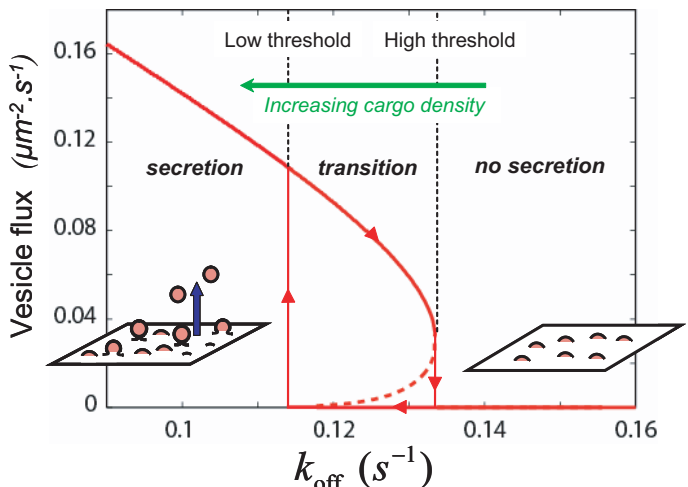}}
\caption{number of secreted vesicle per $\mu m^2.s$ ($=j_{\rm v}\times 1\mu m/s_0$) 
as a function of the rate of coatomer desorption from the membrane $k_{\rm{off}}$. 
The transition to vesiculation is discontinuous, and is characterized by an hysteretic cycle (arrows), 
with high and low turnover thresholds. The parameters are the same as those used for Fig.2 and 3.}
\end{figure}

\subsection{Regulation of vesicle secretion by cargo} 
The adsorption flux $J_{\rm{on}}$, and the desorption rate $k_{\rm{off}}$, 
of coat components at the membrane tightly control vesicle secretion. 
Recent fluorescence  experiments on COPs coat suggest that these rates vary with 
the amount of cargo present at the membrane.
For COPI, the presence of extra cargo leads to significant increase of the amount of 
coat components at the Golgi membrane, which could reflect
either the increase of $J_{\rm{on}}$ or the decrease of $k_{\rm{off}}$ \cite{Lip05}. 
For COPII, FRAP experiments show that the coatomer
exchange rate between the ER membrane and the cytosol is doubled in the absence of cargo. 
This has been attributed to the increase of $k_{\rm{off}}$ with decreasing cargo density \cite{Pep06}.

Our model predicts that vesicles can only be secreted if the recycling rate $k_{\rm off}$ 
is below a critical value (Fig.4). By increasing the lifetime of the coatomers at the membrane, 
the presence of cargo is thus expected to promote vesicle secretion, and a
minimal amount of cargo at the membrane might actually be {\em required} for transport vesicle to be secreted. 
The two, high and low, recycling
thresholds of Fig.4 would then corresponds to two critical cargo densities (low and high, respectively). 

Considering that newly synthesized
cargo is brought to the membrane at a (slow) steady rate and is removed by vesiculation, membrane-bound  
cargo accumulates in the no-secretion regime, thereby
decreasing $k_{\rm off}$ and moving the system toward the secretion regime. 
Above the high cargo density threshold, the coat-machinery abruptly escapes the stationary 
pits regime and switches to vesicle production (Fig.4). This results in a decrease of membrane-bound cargo, 
which increases $k_{\rm off}$ and moves the system toward the quiescent state. Below the low cargo density 
threshold, vesicle secretion is switched off, letting the cargo accumulate until the high-density threshold 
is reached and vesicle production is resumed, starting a new cycle. Under constant cargo in-flux, 
the system should thus periodically switch between quiescent phases and phases of vesicle secretion, following the hysteretic loop of Fig.4.
If cargo synthesis is irregular,  the membrane waits for sufficient accumulation of cargo
between transient residences in the secreting state, where the accumulated cargo is released. 
The vesicle flux over time should then appear as an
irregular pulsed signal. Recording the total vesicle out-flux over an extended patch of secretory
membrane over time should thus be of high interest. Oscillatory or pulsed vesicle secretion
would be a signature of the {\em secretory switch} uncovered by our
theoretical analysis (within our framework, steady secretion would indicate that cargo synthesis is sufficiently fast to 
compensate the secreted cargo).

This accumulator mechanism would provide functional efficiency to the
secretory membrane, as it would prevent the futile delivery of empty
vesicles. Strikingly, analysis of COPI vesicles in mutant cells where
arf1 is unable to hydrolyse GTP have revealed a much lower cargo
content than in normal cells \cite{Pep00}. This observation supports our
prediction. Indeed, in the absence of GTP hydrolysis ($k_{\rm{off}}=0$), the switch is gone and vesicle secretion remains  ``on'' regardless of the available  amount of cargo.

The existence of an hysteretic cycle can also positively impact on the  rate of cargo delivery. At steady state, the flux of cargo delivered by a secretory membrane that would not possesses an unstable regime would automatically adjust to the flux of synthesized cargo. Here, and while the system is traveling along the secretion branch of the hysteretic cycle, the flux of secreted cargo is mainly controlled by the kinetics of coat formation, and can potentially be much larger than the rate of cargo synthesis.


\subsection{Effects of the membrane tension: Regulation of vesicle formation
and coat flattening}

Our calculation has shown  that membrane tension plays essentially the same role 
as coatomer recycling in opposing vesicle secretion (\eq{El}). Vesiculation is only possible below a high tension threshold, and the oscillatory behavior described above may also result from variations of tension. Secretion removes membrane
area from the organelle and may increase its tension, while the fusion of incoming vesicles \cite{Bas06} or other
regulatory mechanisms \cite{Sen06}, dynamically relaxe tension. Vesicle secretion would thus occur only when enough
membrane area has been accumulated to relax the tension. Such a
mechanism  suggests a coordination of purely mechanical origin between absorption and
release of vesicles, and could prevent the uncontroled shrinking of secretory
compartments. Remarkably, the Golgi strikingly crumbles in cells
where the GTP hydrolysis in COPI-coat is rendered inoperant \cite{Pep00}.

In \eq{El0}, the effect of membrane tension was computed assuming that
the coat rigidity $\kappa$ (unit of energy) was sufficiently strong
to impose the domain curvature, which remained constant regardless of
the coat size and the membrane properties. This simplification ceases
to be valid under high membrane tension ($\sigma\sim\kappa/R_0^2$,
where $R_0$ is the radius of curvature of a tensionless coat). Higher tensions result in a flattening of the coat (see SI, section II.B). Beyond  a tension threshold $\sigma>\kappa/(2 R_0^2)$  ($\simeq 10^{-4}J/{\text m}^2$ for the coat rigidity $\kappa=100\kT$), the formation of a
closed sphere becomes impossible and protein aggregates should grow as flat patches. High membrane tension would thus
favor the formation of flat coatomer aggregates which size is
limited by coatomer recycling \cite{turner_raft}. Such "flat lattice" are indeed observed
for Clathrin coats at the basal membrane of adhered cells \cite{Lam07}, where adhesive proteins are expected to generate high membrane tensions.


\subsection{Concluding remarks}

The model presented here is the simplest implementation of a kinetic model of coated vesicle formation where coat growth competes with the inactivation of coat components. The {\em secretory switch} revealed by our work is very robust and relies solely on the (non-equilibrium) Thermodynamics of coat formation. Beyond the qualitative agreement with experimental findings discussed in the previous section, we hope that future experiments can further test some of our predictions, which include: 
{\em i}) the existence of metastable domains of intermediate size,
{\em ii})  the role of membrane tension in preventing the formation of curved protein coat, and eventually its involvement in the formation of flat lattices,
{\em iii}) the oscillatory or pulsed secretion of vesicles in time.

In this study, we have used the crudest possible description of the coat structure, thus avoiding to deal with structural details of specific protein coats. Further experimental observation, {\em e.g} on biomimetic systems, could motivate the building of models focusing on the dynamics of a single coat.
Supplementary degrees of freedom for the coat shape could be considered, allowing for the competitive growth of structures of various morphologies (tubules \cite{Ant06}, spherical caps and flat lattices \cite{Lam07}), observed in living cells and biomimetic systems.

In the same spirit, the heterogeneties of the coat structures
may be included in the model.
The coupling beween the GTP hydrolysis rate and
the curvature (COPI) or the degree of polymerization (COPII) revealed
in recent experiments \cite{Ant03,Ant05}, suggest that
monomers inactivation may be enhanced at the coat center
and limited at the boundary.
One could then imagine the formation of layer of active monomers
preventing coat disassembly \cite{Ant03,Ant05}.
Such properties suggest a strong analogy with microtubules \cite{Lip03}, and it is tempting to imagine exotic growth dynamics with shrinking
cascades such as those observed for microtubules \cite{Lei93}.


\begin{acknowledgments}
We thanks Jean-Baptiste Manneville (Institut Curie-Paris) for very stimulating discussions and critical reading of the manuscript.
\end{acknowledgments}

\end{article}


\begin{thebibliography}{20}

\bibitem{Heu01} Matsuoka, K., Schekman, R., Orci, L. \& Heuser, J.E.
(2001) Surface structure of the COPII-coated vesicle.
{\it Proc. Natl. Acad. Sci USA} {\bf 98}, 13705-13709.

\bibitem{Kir01} Lederkremer, G.Z., Cheng, Y. Petre, B.M., Vogan, E.,
Springer, S.,
Schekman, R., Waltz, T. \& Kirchhausen, T. (2001) 
Structure of the Sec23p/24p and Sec13p/31p complexes of COPII.
{\it Proc. Natl. Acad. Sci USA} {\bf 98}, 10704-10709.

\bibitem{Bar02} Barlowe, C. (2002) 
COPII-dependent transport from the endoplasmic reticulum.
{\it Current Opinion in Cell Biology} {\bf 14}, 417-422.

\bibitem{Mills04} Mc Mahon, H.T. \& Mills, I.G. (2004) 
COP and clathrin-coated vesicle budding: different pathways, common approaches.
{\it Current Opinion in Cell Biology} {\bf 16}, 379-391.

\bibitem{Gal05} Mc Mahon, H.T. \& Gallop, J.L.  (2005) 
Membrane curvature and mechanisms of dynamic cell membrane remodelling.
{\it Nature} {\bf 438}, 590-595.

\bibitem{Ant06} Antonny, B. (2006) 
Membrane deformation by protein coat.
{\it Current Opinion in Cell Biology} {\bf 18}, 386-394.

\bibitem{Yeu98} Matsuoka, K., Orci, L., Amherdt, M., Bednarek, S. Y., 
Hamamoto, S., Schekman, R.\& Yeung, T. (1998) 
COPII-Coated Vesicle Formation Reconstituted with Purified Coat Proteins and Chemically Defined Liposomes. 
{\it Cell.} {\bf 93}, 253-275.


\bibitem{Orc98} Spang, A., Matsuoka, K., Hamamoto, S., Schekman, R.\&
Orci, L. (1998) 
Coatomer, Arf1p, and nucleotide are required to bud coat protein complex I-coated vesicles from large synthetic liposomes.
{\it Proc. Natl. Acad. Sci USA} {\bf 95}, 11199-11204.

\bibitem{Cam98} Takei, K., Haucke, V., Slepnev, V., Farsad, K.,
Salazar, M. Chen, H. \& De Camilli, P .(1998) 
Generation of Coated Intermediates of Clathrin-Mediated Endocytosis on Protein-Free Liposomes. 
{\it Cell} {\bf 94}, 131-141.

\bibitem{Lip01} Presley, J.F., Ward, T.H., Pfeifer, A.C., Siggia, E.,
Phair, R.D. \& Lippincott-Schwartz, J. (2002) 
Dissection of COPI and Arf1 dynamics \textit{in vivo} and role in Golgi membrane transport.
{\it Nature} {\bf 417}, 187-193.

\bibitem{Lip03} Lippincott-Schwartz, J. \& Liu, W. (2003) 
Coat control by curvature.
{\it Nature} {\bf 426}, 507-508.

\bibitem{Lip05} Liu, W., Duden, R., Phair, R.D. \& Lippincott-
Schwartz, J. (2005) 
ArfGAP1 dynamics and its role in COPI coat assembly on Golgi membranes of living cells.
{\it J. Cell Bio.} {\bf 168}, 1053-1063.

\bibitem{Ung06} Hinrichsen, L., Meyerholz, A., Groos, S. \&
Ungewickell, E.J. (2006) 
Bending a membrane: How clathrin affects budding.
{\it Proc. Natl. Acad. Sci. USA} {\bf 103}, 8715-8720.

\bibitem{Ant03} Bigay, J., Gounon, P., Robineau, S. \& Antonny, B.
(2003) 
Lipid packing sensed by ArfGAP1 couples COPI coat disassembly to membrane bilayer curvature.
{\it Nature} {\bf 426}, 563-566.


\bibitem{Pep06} Forster, R., Weiss, M., Zimmermann, T., Reynaud, E.G.,
Verissimo, F., Stephens, D.J. \& Pepperkok, R. (2006) 
Secretory Cargo Regulates the Turnover of COPII Subunits at Single ER Exit Sites. 
{\it Curr. Biol.} {\bf 16}, 173-179.

\bibitem{sens_cav} Sens, P. \& Turner, M.S. (2004) 
Theoretical Model for the Formation of Caveolae and Similar Membrane Invaginations. 
{\it Biophys. J.}
{\bf 86}, 1-9



\bibitem{tension_ref} Sheetz M. P. (2001)  Cell control by membrane?cytoskeleton adhesion {\em Nat. Rev. Mol. Cell Biol.} {\bf 2}, 392-395, Upadhyaya, A. \& and Sheetz, M.P. (2004) Tension in Tubulovescilar Networks of Golgi and Endoplasmic Reticulum Membranes. {\em Biophys. J.} {\bf 86}, 2923-2928

\bibitem{Sen06} Sens, P. \& Turner, M. (2006) 
Budded membrane microdomains as tension regulators. 
{\it Phys. Rev. E} {\bf 73}, 031918.


\bibitem{isra} Israelachvili J. (1991) {\it Intermolecular \& Surface Forces}, 2nd ed., 
Academic Press, San Diego

\bibitem{Pep00} Pepperkok, R., Whitney, J.A., Gomez, M. \& Kreis,
T.E. (2000) 
COPI vesicles accumulating in the presence of a GTP restricted Arf1 mutant
are depleted of anterograde and retrograde cargo. 
{\it J. Cell Science} {\bf 113}, 135-144.


\bibitem{Bas06} Solon, J., Preceaux, J., Girard, P., Faur\'e, M.-C.,
Prost, J. \& Bassereau, P. (2006) 
Negative Tension Induced by Lipid Uptake. 
{\it Phys. Rev. Lett.} {\bf 97}, 098103.



\bibitem{turner_raft} Matthew S. Turner, M. S., Sens, P. \& Socci, N. D. (2004) 
Nonequilibrium Raftlike Membrane Domains under Continuous Recycling. 
{\em Phys. Rev. Let.} {\bf 95}, 168301 

\bibitem{Lam07} Benmerah, A. \& Lamaze, C. (2007) 
Clathrin Coated Pits: Vive la Difference? {\it Traffic} {\bf
8}, 970-982



\bibitem{Ant05} Antonny, B., Bigay, J., Casella, J.-F., Drin, G.,
Mesmin, B. \& Gounon, P. (2005) 
Membrane curvature and the control of GTP hydrolysis in Arf1 during COPI vesicle formation. 
{\it Biochemical Society Transactions} {\bf 33}, 619-622.

\bibitem{Lei93} Dogterom, M. \& Leibler, S. (1993) 
Physical aspects of the growth and regulation of microtubule structures. 
{\it Phys. Rev. Lett.} {\bf 70}, 1347-1350.





\end{thebibliography}
\end{document}